\documentstyle[12pt]{article}

\newcommand{\be}{\begin{equation}}
\newcommand{\ee}{\end{equation}}
\newcommand{\rr}{\rho_\nu (\stackrel{\rightarrow}{r})}
\newcommand{\rs}{\stackrel{-}{\rho} (\stackrel{\rightarrow}{r})}
\newcommand{\ppd}{\stackrel{\dagger}{\psi}_\nu (\stackrel{\rightarrow}{r})}
\newcommand{\pp}{\psi_\nu (\stackrel{\rightarrow}{r})}

\begin{document}

\begin{center}
{\bf CHAOTIC LATTICE - GAS MODEL \\ [5mm]
V.I.Yukalov and E.P.Yukalova} \\ [3mm]
{\it Department of Mathematics and Statistics \\
Queen's University, Kingston \\
Ontario K7L 3N6, Canada}
\end{center}

\vspace{10cm}

{\bf PACS:} 0520; 0540; 0545

{\bf keywords:} statistical mechanics, fluctuation phenomena,

\hspace{2.1cm} chaotic systems, lattice - gas model

\newpage 

\begin{abstract}

A nonuniform system is considered consisting of two phases with different 
densities of particles. At each given time the distribution of the phases 
in space is chaotic: each phase filling a set of regions with random shapes 
and locations. A chaotic diffusion process intermixes these regions by varying
their shapes and locations in a random way. To investigate the statistical 
properties of such a system, it is exemplified by a lattice--gas model. 
Conditions are analysed when this chaotic lattice--gas model can become 
thermodynamically more stable than the usual model describing a pure 
one--phase system. 
\end{abstract}

\newpage

\section{Introduction}

According to the space distribution of particles, there are three types of 
stable matter: uniform, periodic and quasiperiodic. The uniform matter is 
presented by gases and fluids. This can be also called plasma if it consists 
of charged particles. Periodic systems are crystals and quasiperiodic ones 
are incommensurate crystals and quasicrystals [1].

Amorphous solids, or glasses, are not stable, but they are metastable. 
Metastable systems, contrary to stable ones, have finite lifetimes, although 
the latter can be quite long, as it is, e.g., for  the window glass. A more 
detailed description of metastable systems can be found in reviews [2,3].

The dependence of system characteristics on space variables can be analysed 
from the point of view of dynamical theory by analogy with the dependence of 
trajectories on time. Basing on this analogy, amorphous solids could be called
turbulent crystals [4]. The breaking of a regular periodicity in crystals by 
defects can be named the defect turbulence [5]. And an irregular distribution 
of particles, which is neither uniform nor periodic or quasiperiodic, is 
called chaotic [6].

As far as amorphous solids, having chaotic distribution of particles in space,
are metastable, their consideration in time is limited by their lifetime. The 
state, existing during a finite time $ \; \tau_f \; $ that is much longer 
than an observation time $ \; \tau_{obs} \; $ is called frozen 
($ \; \tau_{obs} \ll \tau_f \; $). A frozen chaotic system, after its 
lifetime, transforms, as time goes to infinity, into either uniform or 
periodic (quasiperiodic) system. This can be equivalently said that a frozen 
chaotic system is thermodynamically unstable.

The question arises whether only uniform, periodic and quasiperiodic space 
structures can be stable, or a spatially chaotic system also can exist an 
infinite time. An analogous question has been put forward by Ruelle [4] who 
has asked whether there can exist a frozen turbulent structure with infinite 
lifetime. What is known now about the real  as well as model systems offers 
the negative answer to the Ruelle question, although this does not prove, of 
course, that such a situation is principially impossible. The question that 
we pose here is more general: we ask whether there can exist a 
thermodynamically stable chaotic system (not necessarily frozen).

Let us concretize this question. Suppose 
$ \; \rho (\stackrel{\rightarrow}{r},t) \; $ is the distribution of particle 
density as a function of the real - space vector, 
$ \; \stackrel{\rightarrow}{r} \; $, and time, $ \; t \; $. Assume that the 
limit
$$ \lim_{t \rightarrow \infty} \rho (\stackrel{\rightarrow}{r},t) = 
\rho (\stackrel{\rightarrow}{r}) $$
exists. If the resulting state of the system with the density 
$ \; \rho (\stackrel{\rightarrow}{r}) \; $ is stable against small 
perturbations and has infinite lifetime, it is called thermodynamically 
stable. One can define a thermodynamic potential enjoying the property of 
convexity and being  minimal for a thermodynamically stable system. The 
latter usually has well defined symmetry properties governing the behavior 
of the density $ \; \rho (\stackrel{\rightarrow}{r}) \; $ as a function of 
space coordinates [1,3,7]. If the system is uniform, then the density is 
translationally invariant with respect to an arbitrary space vector 
$ \; \stackrel{\rightarrow}{r} \in {\bf R}^3 \; $, that is
$$ \rho ( \stackrel{\rightarrow}{r} ) = \rho (0).  \qquad 
({\it uniformity}) $$
A periodic structure, or a crystal, is described by a basis 
$ \; \{ \stackrel{\rightarrow}{b}_\alpha | \; \alpha = 1,2,3 \} \; $ of 
generating lattice vectors $ \; \stackrel{\rightarrow}{b}_\alpha \; $, 
with respect to which the density is periodic,
$$ \rho (\stackrel{\rightarrow}{r}) = \rho (\stackrel{\rightarrow}{r} +
\stackrel{\rightarrow}{b}_\alpha ) , \qquad ({\it periodicity}) $$
implying that $ \; \rho (\stackrel{\rightarrow}{r} ) \; $ is not constant.

A quasiperiodic structure is characterized by a basis
$ \; \{ \stackrel{\rightarrow}{b}_\alpha |\; \alpha = 1,2,\ldots , p ; 
\\ p > 3 \} \; $ of integrally independent vectors 
$ \; \stackrel{\rightarrow}{b}_\alpha \; $ such that the density can be 
expressed as a $ \; p \; $- fold Fourier series
$$ \rho (\stackrel{\rightarrow}{r}) = 
\sum_{n_1\ldots n_p} \rho_{n_1\ldots n_p}
e^{i(n_1\stackrel{\rightarrow}{g}_1 + n_2\stackrel{\rightarrow}{g}_2 +
 \ldots + n_p\stackrel{\rightarrow}{g}_p)\stackrel{\rightarrow}{r}} , 
\qquad  ({\it quasiperiodicity}) $$
in which $ \; n_\alpha \; $ are integers and the set 
$ \; \{ \stackrel{\rightarrow}{g}_\alpha \} \; $ is the reciprocal basis 
defined by the equations
$$ \stackrel{\rightarrow}{g}_\alpha\stackrel{\rightarrow}{b}_\alpha = 2\pi
\qquad (\alpha = 1,2,\ldots , p > 3 ) . $$
The integral independence of $ \; \stackrel{\rightarrow}{b}_\alpha \; $ means 
that the linear combination
$$ n_1\stackrel{\rightarrow}{b}_1 + n_2\stackrel{\rightarrow}{b}_2 + \ldots +
n_p\stackrel{\rightarrow}{b}_p \neq 0 \; $$
is nonzero for any nontrivial $ \; p \; $- tuple of integers 
$ \; (n_1,n_2,\ldots , n_p) \; $, i.e. when at least one of $ \; n_\alpha \; $
is nonzero. Note that the structure is quasiperiodic if and only if the rank 
$ \; p \; $ of the basis $ \; \{ \stackrel{\rightarrow}{b}_\alpha \} \; $ is 
greater than the space dimensionality: $ \; p > 3 \; $. Although, probably, 
the majority of quasiperiodic structures are metastable, they also can indeed 
form stable equilibrium states [8], with free energies lower than those of the
conventional crystalline structures. Recall that if 
$ \; \rho (\stackrel{\rightarrow}{r}) \; $ is chaotic in real space, as it is 
for different glasses, e.g. metallic glasses, vitreous silica or amorphous 
polymers [9,10], then such a system is always thermodynamically metastable.

Suppose now that the particle density does not tend to a stationary limit as 
time tends to infinity,
$$ \rho (\stackrel{\rightarrow}{r},t) \rightarrow 
\rho_c(\stackrel{\rightarrow}{r},t) ; \qquad t \rightarrow \infty , $$
but displays spatiotemporal chaos [5]. The average density of the whole 
system is assumed to be fixed,
$$ \frac{1}{V}\int \rho (\stackrel{\rightarrow}{r},t) 
d\stackrel{\rightarrow}{r}
= \frac{1}{V}\int\rho_c(\stackrel{\rightarrow}{r},t) 
d\stackrel{\rightarrow}{r} = \rho ; $$
here $ \; V \; $ is the system volume. A function 
$ \; \rho (\stackrel{\rightarrow}{r},t) \; $ for a given 
$ \; \stackrel{\rightarrow}{r} \; $, exhibits temporal chaos if its time 
motion is locally unstable. We may say that the motion is locally unstable, 
or locally chaotic, if a local Lyapunov exponent is positive:
$$ \lambda ( \stackrel{\rightarrow}{r}) \equiv \lim_{t \rightarrow \infty}
\lim_{\delta\rho (\stackrel{\rightarrow}{r},t_0) \rightarrow 0} \frac{1}{t}
\ln \left | 
\frac{\delta\rho (\stackrel{\rightarrow}{r},t_0+t)}{\delta\rho ( 
\stackrel{\rightarrow}{r},t_0) } \right | > 0 . $$

The case when $ \; \rho (\stackrel{\rightarrow}{r},t) \; $, under fixed 
$ \; \stackrel{\rightarrow}{r} \; $, tends to a time - independent limit 
$ \; \rho (\stackrel{\rightarrow}{r}) \; $, as $ \; t \rightarrow \infty \; $,
describes the passage of a system to an equilibrium state. In the language of 
dynamical theory such a state corresponds to an attractor of fixed - point 
type [3]. Thus, the equilibrium uniform, spatially periodic and spatially 
quasiperiodic states are fixpoint attractors.

The situation when a system tends, as time increases, to a chaotic state, may 
be associated with the existence of a chaotic attractor. In a system with 
dissipation the chaotic attractor is fractal, or strange [11]. As far as the 
equations of motion for $ \;  \rho (\stackrel{\rightarrow}{r},t) \; $ are 
partial differential equations involving derivatives with respect to space 
variables as well as with respect to time, we have to speak about 
spatiotemporal chaos. Such evolution equations correspond to what 
is called infinite - dimensional dynamical systems [12-14].

A chaotic state can be averaged with respect to time. For example, for the 
particle density we may define
$$  \stackrel{-}{\rho} (\stackrel{\rightarrow}{r}) \equiv \lim_{\tau 
\rightarrow \infty} \frac{1}{\tau} \int_{0}^{\tau} 
\rho (\stackrel{\rightarrow}{r},t)dt . $$
The average state, independent of time, can be put into correspondence with 
an effective equilibrium state, or the state equilibrium on the average [3], 
for which an effective free energy can also be defined. When the latter is 
lower than the free energy of any equilibrium state, whether uniform, 
spatially periodic or spatially quasiperiodic, then such a state is 
thermodynamically, or globally, stable, although dynamically, or locally, it 
is unstable.

The question we suggest in this paper is as follows: {\it Can a statistical 
system with spatiotemporal chaos be globally stable}? We construct and analyse
a model giving the positive answer to this question.

\section{Two - density system}

Before considering a particular realization of a chaotic statistical system it
is worth paying some attention to terminology. Since we are dealing not with 
simple dynamical systems but with statistical systems containing infinitely 
many degrees of freedom, we can call the chaos appearing in the latter the 
{\it statistical chaos}. This chaotic state evolves in a system automatically,
without any special fine - tuning of external fields; hence this can be named 
the {\it self - organized chaos}. The time duration of chaotic fluctuations 
and their spatial sizes are distributed in a quite large scale, somewhat 
similar to microscopic critical fluctuations of equilibrium systems, thus, 
one could use the term the {\it critical chaos}. This should be distinguished
 from the self - organized criticality [15] which is a generalization of the 
equilibrium critical state to nonequilibrium systems. The main feature 
distinguishing the chaos we consider is that, although each chaotic 
fluctuation can have any duration, size, and shape, but there can be 
defined, nevertheless, their characteristic, or average, lifetime, 
$ \; \tau_f \; $, and linear size, $ \; l_f \; $, satisfying the inequalities
$$ \tau_{loc} \ll \tau_f \ll \tau_{obs}, \qquad a_0 \ll l_f \ll L , $$
in which $ \; \tau_{loc} \; $ is called the local equilibrium time; 
$ \; \tau_{obs} \; $, the observation time, $ \; a_0 \; $ is the average 
distance between the objects composing the system, and $ \; L \; $ is a linear
size of the latter [3]. Because of this intermediate, or mesoscopic, 
character, it is logical to title the considered phenomenon the 
{\it mesoscopic chaos}. One of the most important observable quantities 
characterizing the properties of s system is the order parameter, generally, 
being a function of space and time variables. The examples are: the particle 
density $ \; \rho (\stackrel{\rightarrow}{r},t) \; $ discussed above, the 
local magnetization, polarization or superconducting gap. In chaotic systems, 
the mesoscopic fluctuations of order parameters are clearly interpreted as 
fluctuating nuclei, or gems, of competing phases. Therefore, such a mesoscopic
chaos can be called as well the {\it heterophase chaos}.

Now, let us return to the concrete case when the chaos in a system is related 
to mesoscopic fluctuations of the particle density 
$ \; \rho (\stackrel{\rightarrow}{r},t) \; $. Suppose that at each fixed time 
the function $ \; \rho (\stackrel{\rightarrow}{r},t) \; $ describes a random 
distribution in space of two regions with different densities, $\;\rho_1\;$ 
and $ \; \rho_2 \; $, such that
\begin{equation}
\rho_1 > \rho_2 \qquad \left ( \rho_\nu \equiv \frac{N_\nu}{V_\nu} \right ) ,
\end{equation}
where $ \; N_\nu \; $ and $ \; V_\nu \; $ are the total number of particles 
and the volume, respectively, corresponding to a $ \; \nu \; $- phase 
($ \; \nu = 1,2 \; $). One can distinguish the regions occupied by different 
phases owing to the mesoscopic nature of the density fluctuations considered. 
The latter are accompanied, of course, by the fluctuations of other physical 
quantities, for example, by rare large energy fluctuations [16,17]. However, 
we shall speak in what follows mainly about the mesoscopic density 
fluctuations since it is just density that plays the role of a natural 
parameter distinguishing different phases.

A physical meaning which we could ascribe to these phases with two different 
densities is by calling the denser phase liquid and the less dense one gas. 
Then, we could imagine that our system is a kind of fog consisting of liquid 
droplets inside gas. Imagining this, we should not identify our chaotic fog 
with the real fog corresponding rather to a frozen droplet configuration, as 
the average lifetime of real droplets is of the order or longer than the 
observation time. As is discussed in the Introduction, such a system with a 
frozen space disorder is metastable, as the real fog is. However, some 
thermodynamic peculiarities of our chaotic fog can be related to the 
so - called droplet singularities [18,19] of liquid - gas systems.

When a persistent chaotic state develops in a system, this means that such a 
state is a chaotic attractor. In dissipative systems, chaotic attractors are 
fractal, or strange [11,12,20]. A chaotic attractor, being metrically 
indecomposable (metrically transitive) allows to define an invariant 
probability measure. The property of metric transitivity for the considered 
case of a system with heterophase chaos is what has been called [3] the 
heterophase quasiergodicity. For a metrically transitive system, time
 averages can be replaced by ensemble averages. To this end, we have to 
introduce a probability measure describing the distribution of phases in 
space. The differential probability measure can be written as 
$$ D\mu (\xi ) = \hat \rho_{loc} (\xi ) D\xi , $$
where $ \; \hat \rho_{loc} (\xi ) \; $ is a locally - equilibrium density 
matrix, and $ \; D\xi \; $ is a functional differential over a set
$$ \xi = \{ \xi_\nu (\stackrel{\rightarrow}{r}) | \; \nu = 1,2,\ldots ;
\stackrel{\rightarrow}{r}  \in {\bf V} \} $$ 
of the manifold indicator functions
\begin{eqnarray}
\xi_\nu (\stackrel{\rightarrow}{r}) = \left \{ \begin{array}{cc}
1 , &  \stackrel{\rightarrow}{r} \in {\bf V}_\nu \\ \nonumber
0,  &  \stackrel{\rightarrow}{r} \not\in {\bf V}_\nu ; \end{array} \right. 
\end{eqnarray}
$ \; {\bf V} \; $ and $ \; {\bf V}_\nu \; $ meaning the whole volume occupied 
by the system, and a region of the volume filled by a $ \; \nu \; $- phase, 
respectively [3]. The averaging over the ensemble of phase configurations 
makes it possible to define an effective phase - replica system that is 
equilibrium on the average [3]. The procedure of overaging over phase 
configurations with all foundations and details has been expounded in a 
series of papers [21-24] and reviewed in [3]. As a result, we can define an 
effective phase - replica Hamiltonian which for the two - phase case takes 
the form
\begin{equation}
\stackrel{-}{H} = H_1 \oplus H_2 ,
\end{equation}
$$ H_\nu = w_\nu \int \stackrel{\dagger}{\psi}_\nu 
(\stackrel{\rightarrow}{r})
\left [ K(\stackrel{\rightarrow}{r}) - \mu \right ] 
\psi_\nu (\stackrel{\rightarrow}{r})d\stackrel{\rightarrow}{r} + $$
$$ \frac{1}{2}w^2_\nu\int  
\stackrel{\dagger}{\psi}_\nu (\stackrel{\rightarrow}{r})
 \stackrel{\dagger}{\psi}_\nu (\stackrel{\rightarrow}{r}')
\Phi (\stackrel{\rightarrow}{r} - \stackrel{\rightarrow}{r}') 
\psi_\nu (\stackrel{\rightarrow}{r}')\psi_\nu (\stackrel{\rightarrow}{r})
d\stackrel{\rightarrow}{r}d\stackrel{\rightarrow}{r}'  , $$
where $ \; K(\stackrel{\rightarrow}{r}) \; $ is the kinetic - energy operator 
including external fields, if any; $ \; \mu \; $, chemical potential; 
$ \; \Phi (\stackrel{\rightarrow}{r} - \stackrel{\rightarrow}{r}') \; $, 
interaction potential; and $ \; \psi_\nu (\stackrel{\rightarrow}{r}) \; $ is a
field - operator representation for the $ \; \nu \; $- phase. The 
renormalizing factors $ \; w_\nu \equiv V_\nu/V \; $ are the geometric 
probabilities of the corresponding phases, satisfying the condition
\begin{equation}
w_1 + w_2 = 1 , \qquad 0 \leq w_\nu \leq 1 ,
\end{equation}
and given by the minimization of the thermodynamic potential
\begin{equation}
y = -\frac{1}{N}\ln Tr e^{-\beta \stackrel{-}{H}} \qquad 
(\beta \Theta \equiv 1) ,
\end{equation}
in which $ \; N \; $ is the total number of particles in the system, and 
$ \; \Theta \; $ is temperature. The total number of particles is
\begin{equation}
N = N_1 + N_2 ; \qquad N_\nu \equiv - 
< \frac{\partial H_\nu}{\partial \mu} > ,
\end{equation}
where $ \; < \ldots > \; $ implies the statistical average with the 
phase - replica Hamiltonian (2). The number of particles in each of the 
phases is
\begin{equation}
N_\nu = w_\nu \int < \stackrel{\dagger}{\psi}_\nu (\stackrel{\rightarrow}{r}) 
\psi_\nu (\stackrel{\rightarrow}{r})> d \stackrel{\rightarrow}{r} .
\end{equation}
The mean density of particles in the system is a linear combination
\begin{equation}
\rho \equiv \frac{N}{V} = w_1\rho_1 + w_2\rho_2  
\end{equation}
of densities
\begin{equation}
\rho_\nu = \frac{1}{V} \int < \stackrel{\dagger}{\psi}_\nu 
(\stackrel{\rightarrow}{r}) 
\psi_\nu (\stackrel{\rightarrow}{r})> d \stackrel{\rightarrow}{r} ,
\end{equation}
which, by definition (1), are different in the two phases.

For a system equilibrium on average we can define all observable quantities 
as standard statistical averages with the Hamiltonian (2). Thus, for the 
internal energy per particle we have
$$ E \equiv \frac{1}{N} \left ( < \stackrel{-}{H} > + \mu N \right ) = 
\frac{1}{N} \left ( < H_1 > + < H_2 > \right ) + \mu , $$
\begin{equation}
< H_\nu > = w_\nu \left ( K_\nu - \mu \frac{\rho_\nu}{\rho}\right ) N +
w_\nu^2B_\nu N ,
\end{equation}
where the notation is used for the mean kinetic energy
\begin{equation}
K_\nu = \frac{1}{N} 
\int < \stackrel{\dagger}{\psi}_\nu (\stackrel{\rightarrow}{r})
K (\stackrel{\rightarrow}{r}) 
\psi_\nu (\stackrel{\rightarrow}{r})> d \stackrel{\rightarrow}{r} ,
\end{equation}
and mean potential energy
\begin{equation}
B_\nu = \frac{1}{2N} 
\int < \stackrel{\dagger}{\psi}_\nu (\stackrel{\rightarrow}{r})
\stackrel{\dagger}{\psi}_\nu (\stackrel{\rightarrow}{r}')
\Phi (\stackrel{\rightarrow}{r} - \stackrel{\rightarrow}{r}')
\psi_\nu (\stackrel{\rightarrow}{r})
\psi_\nu (\stackrel{\rightarrow}{r}') > d \stackrel{\rightarrow}{r} 
\end{equation}
per particle in each of the phases.

Recall that speaking here about the system equilibrium on average we imply 
an effective system obtained after averaging over phase configurations which 
is equivalent to the averaging with respect to time. The actual system 
describing chaotic motion of two phases is, of course, nonequilibrium [3].

To define the phase probabilities, $ \; w_\nu \; $, providing the minimum 
for the thermodynamic potential (4), under the normalization condition (3), 
it is convenient to use the notation
\begin{equation}
w \equiv w_1 , \qquad w_2 \equiv 1 - w .
\end{equation}
Then, the extremum of (4) with respect to $ \; w \; $ implies
\begin{equation}
\frac{\partial y}{\partial w} = \frac{1}{N} 
< \frac{\partial \stackrel{-}{H}}{\partial w} > = 0 .
\end{equation}
This, with the Hamiltonian (2), yields
\begin{equation}
w = \frac{2B_2 + K_2 - K_1 - (\mu /\rho )( \rho_2 - \rho_1 )}{2(B_1 + B_2)} .
\end{equation}
Equation (14), in compliance with notation (12) and assumption (1), defines 
the probability of the dense phase. The probability of the diluted phase is 
$ \; w_2 = 1 - w \; $.

The phase probabilities given by (12) and (14) provide the minimum of (4) if
\begin{equation}
\frac{\partial^2 y}{\partial w^2} > 0 .
\end{equation}
The latter inequality is equivalent to
\begin{equation}
\left \{ < \frac{\partial^2 \stackrel{-}{H}}{\partial w^2} > - \frac{1}{T} 
< \left ( \frac{\partial \stackrel{-}{H}}{\partial w}\right )^2 >
\right \} > 0 .
\end{equation}
Conditions (15), or (16), are {\it conditions of heterophase stability}. The second term in (16) is always non - negative, thence we can write a simplified version of (16),
\begin{equation}
\left \{ < \frac{\partial^2 \stackrel{-}{H}}{\partial w^2} > \right \} > 0 ,
\end{equation}
which is a {\it necessary conditions of heterophase stability}. The latter, 
taking account of (2), yields
\begin{equation}
B_1 + B_2 > 0 .
\end{equation}
In addition, the second property in (3) gives
\begin{equation}
\frac{\mu}{\rho} \left ( \rho_1 - \rho_2 \right ) -2B_1 \leq K_1 - K_2 
\leq 2B_2 + \frac{\mu}{\rho} \left ( \rho_1 - \rho_2 \right ) .
\end{equation}

Besides, it is worth checking that the system satisfies the usual conditions 
of thermodynamic stability, according to which the specific heat and 
isothermal compressibily are to be positive:
$$ C_\nu = -\Theta \frac{\partial^2 f}{\partial \Theta^2} > 0 , $$
\begin{equation}
\kappa_T = \left ( \rho \frac{\partial P}{\partial \rho}\right )^{-1} > 0 ,
\end{equation}
 here the free energy per particle, $ \; f \; $, and the pressure, 
$ \; P \; $, are related to the thermodynamical potential (4) by the 
equations
$$ f = y\Theta + \mu , \qquad P = \rho^2\frac{\partial f}{\partial \rho} . $$
The system is called absolutely stable in the thermodynamic sense if its 
thermodynamic potential is the lowest possible and all stability conditions 
described above are valid. For brevity, we may call this type of absolute 
thermodynamic stability the {\it global stability}, as compared with the 
local dynamic instability of chaotic motion. The heterophase chaotic system 
is dynamically (locally) unstable, while it can be thermodynamically 
(globally) stable.

\section{Lattice - gas model}

To get more details of statistical properties for a system with heterophase 
chaos, let us pass to an approximation leading to what is termed lattice - gas
model. To this end, we divide the whole volume of the system, $ \; V \; $, 
into $ \; N_L \; $ lattice cells, the density of lattice sites being
\begin{equation}
\rho_0 \equiv \frac{N_L}{V} .
\end{equation}
This quantity can be used for defining the dimensionless densities 
\begin{equation}
n_\nu \equiv \frac{\rho_\nu}{\rho_0} , \qquad n \equiv \frac{\rho}{\rho_0} =
\frac{N}{N_L} .
\end{equation}
With notation (22), we can recast (7) to the equation
\begin{equation}
n = w_1n_1 + w_2n_2 , 
\end{equation}
which may be used for defining the chemical potential $ \; \mu \; $ as a 
function of density and temperature, $ \; \mu (n,\Theta ) \; $.

Introduce a set $\;\{ \stackrel{\rightarrow}{a}_i |\; i=1,2,\ldots, N_L\}\;$ 
of vectors corresponding to lattice sites and define the Wannier functions 
$\;\varphi_n (\stackrel{\rightarrow}{r} -\stackrel{\rightarrow}{a}_i )\;$ 
with $ \; n \; $ indexing quantum states. The field operator 
$ \; \psi_\nu (\stackrel{\rightarrow}{r} ) \; $ can be written as an 
expansion
\begin{equation}
\psi_\nu (\stackrel{\rightarrow}{r}) = 
\sum_{i=1}^{N_L} \sum_{n} c_{in\nu} e_{i\nu}\varphi_n ( 
\stackrel{\rightarrow}{r} - \stackrel{\rightarrow}{a}_i ) , 
\end{equation}
in which $ \; e_{i\nu} = 1,0 \; $ is a variable telling whether the 
$ \; i \; $- lattice cell inside the $ \; \nu \; $- phase is occupied by a 
particle or empty, respectively. Assume that each cell can accept not 
more than one particle, which results in the homeopolarity condition
\begin{equation}
\sum_{n} \stackrel{\dagger}{c}_{in\nu}c_{in\nu} = 1 .
\end{equation}
A standard supposition for a lattice - gas approximation [25] is that 
intercell and interlevel transitions are prohibited. This can be expressed 
in two ways. One possibility [26] is to require the diagonality of the matrix 
elements
$$ < mi | K | nj > = \delta_{mn}\delta_{ij}K_i , $$
$$ < mi,m'i'|\Phi | n'j',nj > = \delta_{mn}\delta_{ij}
\delta_{m'n'}\delta_{i'j'}\Phi_{ij} . $$
Equivalently, we can define the operators $ c_{in\nu} \; $ on a restricted 
space of states with the restriction
\begin{equation}
\stackrel{\dagger}{c}_{im\nu}c_{jn\nu} = 
\delta_{ij}\delta_{mn}  \stackrel{\dagger}{c}_{in\nu}c_{jn\nu} .
\end{equation}
Then, (24), (25) and (26) permit to cast each term of (2) into the form
\begin{equation}
H_\nu = w_\nu \sum_{i=1}^{N_L} (K_i - \mu )e_{i\nu} + \frac{1}{2} 
w_\nu^2 \sum_{i \neq j}^{N_L} \Phi_{ij}e_{i\nu}e_{j\nu} .
\end{equation}
Thus, we obtain the effective phase - replica Hamiltonian 
$ \; \stackrel{-}{H} = H_1 \oplus H_2 \; $, with terms (27), describing the 
statistical properties of a chaotic lattice - gas model.

In the considered case of heterophase chaos, the system consists of two 
randomly distributed phases with different densities. The latter are given 
by (8) which for the lattice gas become
\begin{equation}
\rho_\nu = \frac{1}{V} \sum_{i=1}^{N_L} < e_{i\nu} > .
\end{equation}
As the dimensionless densities (22) we get
\begin{equation}
n_\nu = \frac{1}{N_L} \sum_{i=1}^{N_L} < e_{i\nu} > .
\end{equation}
From the occupation - of - cell variable $ \; e_{i\nu} = 1,0 \; $ one can 
pass to the spin variable $ \; \sigma_{i\nu} = \pm 1 \; $ by using the 
transformation
\begin{equation}
e_{i\nu} = \frac{1}{2} \left ( 1 + \sigma_{i\nu} \right ) , \qquad 
\sigma_{i\nu} = 2e_{i\nu} - 1 .
\end{equation}
Then, for the density (29) we have  
\begin{equation}
n_\nu = \frac{1}{2} \left ( 1 + m_\nu \right ) ,
\end{equation}
where
\begin{equation}
m_\nu = \frac{1}{N_L} \sum_{i=1}^{N_L} < \sigma_{i\nu} > .
\end{equation}
According to (1), the first phase is dense and the second one is diluted, 
which for (31) and (32) means that
\begin{equation}
n_1 > n_2 , \qquad m_1 > m_2 .
\end{equation}
In the quasispin representation the Hamiltonian (27) becomes
\begin{equation}
H_\nu = U_\nu + \frac{1}{8} w_\nu^2 \sum_{i \neq j}^{N_L} \Phi_{ij} 
\sigma_{i\nu} \sigma_{j\nu} - \frac{1}{2}w_\nu \sum_{i=1}^{N_L} 
h_{i\nu}\sigma_{i\nu} ,
\end{equation}
with the notation
$$ U_\nu = \frac{1}{2} N_L w_\nu \left ( K - \mu + 
\frac{1}{4} w_\nu \Phi \right ) , $$
$$ K \equiv \frac{1}{N_L} \sum_{i=1}^{N_L} K_i , \qquad 
\Phi \equiv \frac{1}{N_L} \sum_{i\neq j}^{N_L} \Phi_{ij} , $$
$$ h_{i\nu} = \mu - K_i - \frac{1}{2} w_\nu \Phi . $$ 

The operator from of (34) corresponds to the Ising model with an external 
field. Such a model, as is known, is difficult to treat. For low 
dimensionality, there exists a numerical transfer - matrix method based on a 
real - space Trotter decomposition of the statistical operator [27]. A 
numerical transfer - matrix study based on finite - size scaling [28] is 
applicable to the Ising model in two and three dimensions. However, our aim 
here is not to present some extra accurate calculations but to clarify the
main physical features of a new model. Therefore, at this stage it would be 
unreasonable to sink into clumsy technical details. Instead, we can think 
that $ \; \Phi_{ij} \; $ is a long - range interaction for which the 
mean - field approximation works well. The latter gives simple expressions 
for correlation functions, such as
$$ < \sigma_{i\nu}\sigma_{j\nu} > = m_\nu^2 \qquad ( i \neq j ) . $$
For arbitrary lattice sites we can write
\begin{equation}
 < \sigma_{i\nu}\sigma_{j\nu} > = \delta_{ij} + ( 1 - \delta_{ij})m_\nu^2 .
\end{equation}
Using (35), we are able to continue the explicit analytical investigation of 
our model.

First of all, let us check the conditions of heterophase stability. For 
quantities (10) and (11) we now have
$$ K_\nu = \frac{K}{2n} \left ( 1 + m_\nu \right ) = K\frac{n_\nu}{n} , $$
\begin{equation}
B_\nu = \frac{\Phi}{8n}\left ( 1 + m_\nu \right )^2 = 
\Phi \frac{n_\nu^2}{2n} .
\end{equation}
As far as in the mean - field approximation
$$ < \left ( \frac{\partial \stackrel{-}{H}}{\partial w} \right )^2 > 
 = < \frac{\partial \stackrel{-}{H}}{\partial w} >^2  = 0 , $$
the conditions of stability (16) and (17) coincide with each other and with 
(18). The latter, taking account of (36), leads to the condition of 
heterophase stability
\begin{equation}
\Phi > 0 .
\end{equation}
This shows that to make the heterophase system stable, the particle 
interactions are to be repulsive on average.

For the free energy we obtain
$$ f = \frac{1}{2n} \left ( K - \mu \right ) + \frac{1}{8n} 
\Phi \sum_{\nu} w_\nu^2 ( 1 - m_\nu^2 ) + \mu - $$
\begin{equation}
- \frac{\Theta}{n} \sum_{\nu} \ln \left [ 2\cosh 
\frac{\Phi (1+m_\nu )w_\nu^2 + 2(K - \mu ) w_\nu}{4 \Theta} \right ] .
\end{equation}
The mean quasispin (32) becomes
\begin{equation}
m_\nu = - \tanh \frac{\Phi w_\nu^2 (1 + m_\nu ) + 
2 w_\nu (K - \mu )}{4\Theta} .
\end{equation}
The equation for the phase probabilities can be obtained either from(13), 
that is from the minimization of (38) with respect to $ \; w \; $, or by 
substituting (36) into (14). In the both cases we get, of course, the same 
equation
\begin{equation}
w_1\Phi ( 1 + m_1)^2 + 2m_1(K - \mu ) = w_2 \Phi ( 1+ m_2 )^2 + 
2m_2 (K - \mu ),
\end{equation}
to which we have to add (23) to define the chemical potential.

\section{Thermodynamic properties}

Now we have in hands all information needed for investigating thermodynamic 
properties of the chaotic lattice - gas model, whose phase probabilities and 
chemical potential are given by equations (40) and (23). For convenience, we 
write down these equation
s in the form
\begin{equation}
w = \frac{\Phi n_2^2 + (K - \mu )(n_2 - n_1 )}{\Phi ( n_1^2 + n_2 ^2 )} ,
\end{equation}
\begin{equation}
n = wn_1 + (1 - w)n_2 .
\end {equation}

First, let us show that the considered chaotic state is thermodynamically 
more profitable than a homophase state. For zero temperature, this can be 
readily shown by comparing the internal energies of the chaotic two - phase 
system and of a homogeneous one - phase system. Thus, substituting (36) into 
(9), we find
$$ E = \frac{K}{n} \left ( w_1n_1 + w_2n_2 \right ) + \frac{\Phi}{2n} 
\left ( w_1^2 n_1^2 + w_2^2n_2^2 \right ) . $$
Invoking (42), we obtain
\begin{equation}
E = K + \Phi\frac{n}{2} - \frac{\Phi}{n} w_1w_2n_1n_2 .
\end{equation}
In the case of a homophase system, when one of the phase probabilities is 
zero, we have
\begin{equation}
E_{hom} = K +\Phi \frac{n}{2} .
\end{equation}
The comparison of (43) and (44) proves that
\begin{equation}
E < E_{hom} \qquad ( \Phi > 0 ).
\end{equation}
At finite temperatures, we have to compare (38) with
$$ f_{hom} = \frac{K-\mu_0}{2n} + \frac{1-n}{2}\Phi + \mu_0 - 
\frac{\Theta}{n} \ln \left ( 2\cosh \frac{\Phi n + K -\mu_0}{2\Theta} 
\right ) , $$ 
where
$$ \mu_0 = K +\Phi n + 2\Theta arctanh (2n - 1) . $$
Numerical analysis shows that there always exists a range of temperatures 
when $ \; f < f_{hom} \; $.

To investigate in more details the temperature behavior of thermodynamic 
characteristics of the chaotic lattice - gas model, and to check its stability
conditions, we proceed in what follows to a numerical investigation. For this 
purpose it is convenient introduce a dimensionless chemical potential and 
temperature,
\begin{equation}
\mu_* \equiv \frac{\mu - K}{\Phi} , \qquad T \equiv \frac{\Theta}{\Phi} .
\end{equation}
Let us also use the notation
\begin{equation}
a \equiv n_1 , \qquad b \equiv n_2 \qquad ( a > b ) .
\end{equation}
Eq.(38), with the help of (46) and (47), can be rewritten as
$$ \frac{f - \mu}{\Phi}n = \frac{a}{2} \left ( 1 - a \right )w^2 + \frac{b}{2}
\left ( 1 - b \right ) \left ( 1 - w \right )^2 - \frac{1}{2}\mu_* - $$
\begin{equation}
- T\ln \left [ 4\cosh \frac{aw^2 -\mu_* w}{2T} \cdot 
\cosh \frac{b(1 - w)^2 - \mu_* (1 - w)}{2T} \right ] .
\end{equation}
The quantities $ \; a \; $ and $ \; b \; $ play the role of order parameters 
distinguishing phases with different densities. The equations for these order 
parameters follow from (39) giving
$$ 2a -1 = -\tanh \frac{aw^2 - \mu_* w}{2T} $$
\begin{equation}
2b -1 = -\tanh \frac{b(1-w)^2 - \mu_* (1 -w)}{2T} .
\end{equation}
And the equations for $ \; w \; $ and $ \; \mu_* \; $ are (41) and (42). From 
(41) we can express the chemical potential 
\begin{equation}
\mu_* = \frac{a^2w - b^2(1 - w)}{a - b} , 
\end{equation}
and from (42), the phase probability
\begin{equation}
w = \frac{n - b}{a-b}  \qquad ( a > b ) .
\end{equation}
Substituting (51) into (50), we get
\begin{equation}
\mu_* = \frac{a^2(n-b) + b^2 (n-a )}{(a-b)^2} .
\end{equation}
Equations (49), with the use of (52), become
$$ 2a = 1 - \tanh \left [ b(n-b) \frac{2ab - n(a+b)}{2T(a-b)^3} \right ] , $$
\begin{equation}
2b = 1 + \tanh \left [ a(n-a) \frac{2ab - n(a+b)}{2T(a-b)^3} \right ] .
\end{equation}
In this way, if we solve (53) with respect to $ \; a > b \; $, then we can 
calculate (51) and (52).

Note that (51), together with the property of the phase probability, 
$ \; 0 < w < 1 \; $, impose the following condition
\begin{equation}
b < n < a
\end{equation}
on the solution of (53). Numerical investigation of (53) shows that the 
solutions $ \; a > b \; $ exist provided that
\begin{equation}
0 < n < 0.323 .
\end{equation}
 These solutions are plotted in fig.1.

Figs. 2 and 3 display the temperature behaviour of the dense - phase 
probability (51) and of the dimensionless chemical potential (52), 
respectively. As we see, the probability of the dense phase diminishes with 
temperature, becoming zero at some $ \; T_c (n) \; $.

The pressure of the system is drawn in fig.4. And figs. 5 and 6 demonstrate 
that the stability conditions (20) for the specific heat and isothermal 
compressibility are satisfied.

To illustrate that the condition of heterophase stability (37) is directly 
related to the conditions of thermodynamic stability (20), we also considered 
the case $ \; \Phi < 0 \; $. The corresponding equations for the order 
parameters can be obtained from (53) after the change 
$ \; \Phi \rightarrow -\Phi \; $. The resulting equations do possess 
nontrivial solutions depicted in figs.7,8 and 9. Even more, there appear 
three branches of these solutions. However, for no one of these branches the 
conditions of thermodynamic stability (20) hold. Thus, the conditions of 
heterophase and thermodynamic stability are either both valid or both 
broken.

\section{Discussion}

We have constructed a heterophase lattice - gas model consisting of two 
phases with different densities, one phase may be called dense; another, 
diluted. The dynamics of the phases is chaotic. However, the system, as a 
whole, can be globally stable.

The density of particles in such a system, after averaging over phase 
configurations, or over time, has the form of a sum
\be
\rs = \sum_{\nu}w_\nu\rr ,
\ee
in which
\be
\rr = < \ppd\pp > .
\ee
The space averages of (56) and (57), i.e.,
\be
\rho =\frac{1}{V}\int \rs d\stackrel{\rightarrow}{r} , \qquad
\rho_\nu =\frac{1}{V}\int \rr  d\stackrel{\rightarrow}{r} ,
\ee
lead to the mean densities (7) and (8), respectively.

For the lattice - gas model considered, the density (56), in accordance with 
(57) and (24), is periodic. In this way, although at each phase configuration,
or at any fixed time, there is no periodicity in the system because of random 
distribution of two phases is space, but after averaging over phase 
configurations, or over time, the averaged density of particles (56) happens 
to be periodic.

A straightforward generalization of the model could be if we would ascribe to 
different phases different lattices. Putting into correspondence to each 
phase $ \; N_{L\nu} \; $ cells with the attached lattice vectors 
$ \; \{ \stackrel{\rightarrow}{a}_{i\nu} \} \; $, we may write, instead of 
(24), the expansion
\be
\pp = \sum_{i=1}^{N_{L_\nu}}\sum_{n}c_{in\nu}e_{i\nu}
\varphi_n (\stackrel{\rightarrow}{r} -\stackrel{\rightarrow}{a}) .
\ee
Then, in place of (57) we would have
\be
\rr = \sum_{i=1}^{N_{L_\nu}}\sum_{n} <e_{i\nu}> 
<\stackrel{\dagger}{c}_{in\nu}c_{in\nu}> 
|\varphi_n (\stackrel{\rightarrow}{r} -
\stackrel{\rightarrow}{a}_{i\nu})|^2 ,
\ee 
which is a periodic function,
$$ \rho_\nu (\stackrel{\rightarrow}{r} + 
\stackrel{\rightarrow}{a}_{i\nu}) = \rr $$
If for a two - phase system the vectors $\;\stackrel{\rightarrow}{a}_{1\nu}\;$
and $\;\stackrel{\rightarrow}{a}_{2\nu}\;$ are integrally independent, as is 
defined in the Introduction, then the sum (56) of two periodic functions (59) 
is a quasiperiodic function. Similarly, we could construct a quasiperiodic 
linear combination (56) composed of an arbitrary number of periodic terms.

So, a system can seem, on average, to be quasiperiodic while actually it is 
chaotic. It is possible to speculate that some incommensurate structures could
be of this kind. Whether an order in a system is frozen or is sustained only 
on average, can be checked by appropriate experiments. For example, the 
M\"ossbauer spectroscopy is very sensitive to structural features, and the 
M\"ossbauer factor displays specific anomalies in the presence of lattice 
instabilities, especially when the latter are connected with mesoscopic 
structural fluctuations as in the considered case [29,30]. Anyway, even if 
such systems would not exist in nature, the model investigated here presents 
an illustration of unexpected possibility when the onset of chaos, that is of 
local instability, makes the system globally more stable. However, we think 
that this possibility is not just a strangeness of an artificial model but 
rather a general phenomenon common for many physical systems, the phenomenon 
called [3] the {\it spontaneous breaking of equilibrium}.

\newpage

\begin{center}
{\bf Figure captions}
\end{center}

\vspace{1cm}

\parindent=0pt
{\bf Fig.1.} 

\parindent=0pt
The dimensionless density of the dense phase (solid curves) and the diluted 
phase (dashed curves) vs. temperature for several mean densities. The curves 
end at the points where the system looses its stability.

\vspace{1cm}

{\bf Fig.2}

\parindent=0pt
The probability of the dense phase as a function of temperature.

\vspace{1cm}

{\bf Fig.3}

\parindent=0pt
The dimensionless chemical potential vs. temperature.

\vspace{1cm}

{\bf Fig.4}

\parindent=0pt
The temperature dependence of pressure.

\vspace{1cm}

{\bf Fig.5}

\parindent=0pt
The specific heat as a function of temperature.

\vspace{1cm}

{\bf Fig.6}

\parindent=0pt
The isotermal compressibility vs. temperature.

\vspace{1cm}

{\bf Fig.7}

\parindent=0pt
The dimensionless density of the dense phase (solid curves) and of the 
diluted phase (dashed curves) vs. temperature in the case of the unstable 
system with the mean density $ \; n = 0.20 \; $. Numbers 1,2 and 3 enumerate 
the branches of solutions.

\vspace{1cm}

{\bf Fig.8}

\parindent=0pt
The same as in fig.7 but for the mean density $ \; n = 0.26 \; $.

\vspace{1cm}

{\bf Fig.9}

\parindent=0pt
The dependence of the densities of the dense phase (solid curve) and of the 
diluted phase (dashed curve) on the mean density in the case of the unstable 
system at fixed temperature $ \; T=0.03 \; $.

\end{document}